\documentclass[letterpaper, 10 pt, conference]{ieeeconf}  

\IEEEoverridecommandlockouts     

\usepackage{graphicx}
\usepackage[caption=false]{subfig}
\usepackage[export]{adjustbox}
\usepackage{hyperref}
\usepackage{tikz}
\def\qq#1{`#1'}%

\newcommand\copyrighttext{%
  \textcopyright 2021 IEEE. Personal use of this material is permitted.
  Permission from IEEE must be obtained for all other uses, in any current or future
  media, including reprinting/republishing this material for advertising or promotional
  purposes, creating new collective works, for resale or redistribution to servers or
  lists, or reuse of any copyrighted component of this work in other works. \\
  \\~
  To appear at ISMR 2021, Nov. 17-19, 2021. \\
  
  }
\newcommand\copyrightnotice{%
\begin{tikzpicture}[remember picture,overlay]
\node[anchor=north, yshift=-50pt] at (current page.north) {\mbox{\parbox{\dimexpr\textwidth-\fboxsep-\fboxrule\relax}{\copyrighttext}}};
\end{tikzpicture}%
}

\title{\LARGE \bf
Deep Learning for Needle Detection in a Cannulation Simulator
}

\author{Jianxin Gao$^{\dag}$, Ju Lin, Irfan Kil, Ravikiran B. Singapogu, and Richard E. Groff 

\thanks{$^{\dag}$ To whom all correspondence should be addressed.\newline Jianxin Gao, Ju Lin, and Richard E. Groff are with the Dept. of Electrical \& Computer Engineering, Clemson University, Clemson, SC-29634-0915 ((jianxig, jul, and regroff)@clemson.edu).\newline Irfan Kil is with Applied Medical Resources Corporation, Rancho Santa Margarita, CA-92688 (irfan.kil@appliedmedical.com).\newline Ravikiran B. Singapogu is with the Dept. of Bioengineering, Clemson University, Clemson, SC-29634 (joseph@clemson.edu).}%

}

\begin{document}

\copyrightnotice
\maketitle
\thispagestyle{empty}
\pagestyle{empty}

\begin{abstract}
Cannulation for hemodialysis is the act of inserting a needle into a surgically created vascular access (e.g., an arteriovenous fistula)  for the purpose of dialysis. The main risk associated with cannulation is infiltration, the puncture of the wall of the vascular access after entry, which can cause medical complications. Simulator-based training allows clinicians to gain cannulation experience without putting patients at risk.  In this paper, we propose to use deep-learning-based techniques for detecting, based on video, whether the needle tip is in or has infiltrated the simulated fistula. Three categories of deep neural networks are investigated in this work: modified pre-trained models based on VGG-16 and ResNet-50, light convolutional neural networks (light CNNs), and convolutional recurrent neural networks (CRNNs).  CRNNs consist of convolutional layers and a long short-term memory (LSTM) layer. A data set of cannulation experiments was collected and analyzed. The results show that both the light CNN (test accuracy: 0.983) and the CRNN (test accuracy: 0.983) achieve better performance than the pre-trained baseline models (test accuracy 0.968 for modified VGG-16 and 0.971 for modified ResNet-50). The CRNN was implemented in real time on commodity hardware for use in the cannulation simulator, and the performance was verified.  Deep-learning video analysis is a viable method for detecting needle state in a low cost cannulation simulator. Our data sets and code are released at \url{https://github.com/axin233/DL_for_Needle_Detection_Cannulation}.
\newline

\end{abstract}

\section{INTRODUCTION}

Hemodialysis is the most popular treatment for end-stage kidney disease in the United States \cite{hemodialysis}. It allows patients' blood to circulate through the dialysis machine, which is used to remove excess fluid and waste products from patients' bodies. To perform hemodialysis, patients' arteriovenous fistulas (AVF) or grafts are cannulated. The cannulation process, however, has a high rate of failure. Studies show that approximately 50\% AVF access has mild infiltration injury \cite{lok2020_failure_rate}. Infiltration is the puncture of the wall of the vascular access after entry. Severe infiltration may lead to medical complications such as fistula thrombosis \cite{infiltration}.



 One approach to improving the cannulation success rate is to provide technology to aid clinicians during cannulation.   For example, ultrasound devices that visualize the fistula are actively being developed for this purpose \cite{ultrasound_1,ultrasound_2}.  An alternative approach to improving cannulation success rate is to improve training for clinicians, for example using a high fidelity simulator.    

Our lab has developed a series of simulators for cannulation training based on various technologies. Since needle insertion is critical for understanding the cannulation process \cite{cannulation_process}, these simulators are designed to support both needle puncture and infiltration detection.  Liu et al. detect needle tip location by putting an infrared (IR) detector inside the needle and an IR emitter in the artificial fistula \cite{using_IR, liu_using_IR_2}. Zhang et al. use electromagnetic sensing to track the needle tip position and determine whether the needle tip has pierced or infiltrated the fistula \cite{using_EM}. In the present work, by comparison, the needle tip is identified by analyzing video frames from a camera located inside the simulated fistula. Although the previous version of our simulator \cite{using_cv}  used image-based techniques for detecting the needle tip, it did not support infiltration detection. Additional methods of detecting needle puncture have been developed for other applications, for example, based on force \cite{saito_puncture_detect_force} and bioimpedence \cite{cheng_puncture_detect_bio-impedance}.  Nevertheless, the image-based technique is particularly well-suited for use in the simulator, since it requires no modification of the needle body and no hardware beyond a USB camera and laptop.    


To the best of our knowledge, this paper is the first work to introduce artificial neural networks for needle puncture and infiltration detection. The neural networks are trained and tested on video data collected from a cannulation simulator \cite{using_cv}. Since the video data consists of sequential images, the architecture of the networks includes convolutional layers and long short-term memory (LSTM) layers. Convolutional neural networks (CNNs) have been proven to have high accuracy for image classification \cite{LeNet,AlexNet}. Furthermore, transfer learning allows pre-trained CNN models to be used for specific tasks \cite{fruit_classification,pavement_distress_detection,pest_detection}. LSTM, a special recurrent neural network, supports remembering temporal information but avoids the long-term dependency problem \cite{LSTM}. Studies show that LSTM is suitable for sequence labeling \cite{LSTM_application_Lin, LSTM_application_Ma}. The combination of CNN and LSTM, named CRNN, has shown great success in many sequential tasks~\cite{julpacket2021,Gorron2019_music_detection,wang2019_text_classification}. CRNN takes advantages of CNNs for local feature extraction and LSTMs for temporal summarization of the extracted features. Since our networks only require video frames for needle detection, the resulting simulator can achieve high detection accuracy but use inexpensive hardware. 

This paper is organized as follows. Section II describes the data set, the four network structures, and the real-time experiments. Section III presents performance results for the four network structures and for the real-time experiments. Finally, conclusions  and future work are in Section IV. 


\section{METHODOLOGY}
The cannulation simulator (Fig. \ref{simulator}) described in \cite{using_cv} is used to collect video data. The videos are then used to generate the data set for training the neural networks. The simulator contains a camera (Logitech C920 HD Pro) and an artificial fistula. The camera is used to record videos inside the fistula. The videos are recorded at 30 fps with 640$\times$480 resolution. The artificial fistula, including a tube (diameter: 14 mm) and a rectangular prism base, is made from silicone rubber (Ecoflex 00-30, Smooth-On Inc). Note that the artificial fistula is surrounded by LED strips to provide consistent lighting, independent of external illumination. For our cannulation simulator, the diameter of the insertion area is 10.5 cm.

\subsection{Generating the data set}\label{section_data_set}

\begin{figure}
\subfloat[\label{camLoc}]{%
      \includegraphics[width=0.45\columnwidth]{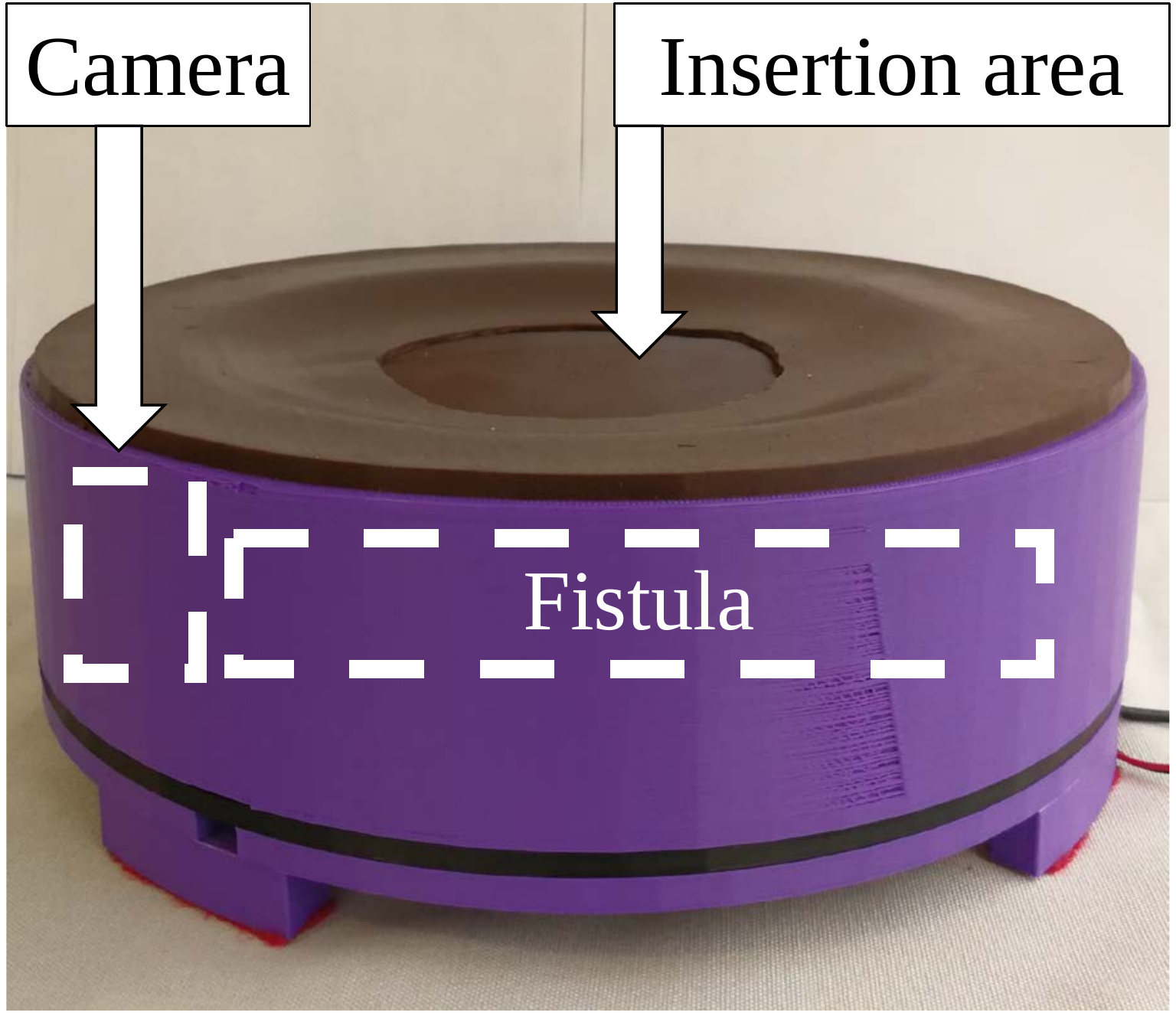}
     }
     \hfill
\subfloat[\label{section}]{%
      \includegraphics[width=0.45\columnwidth]{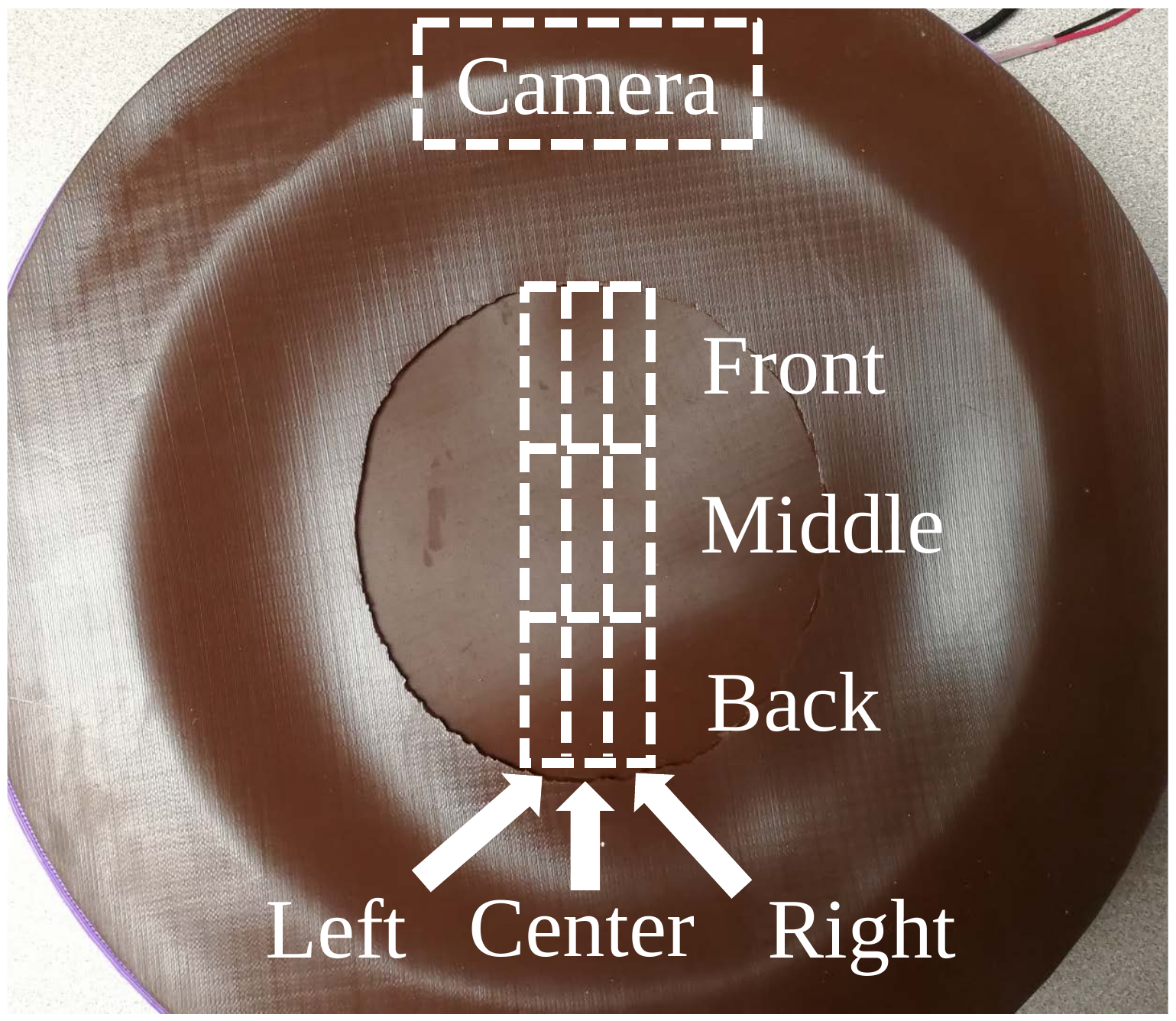}
     }
     \caption{Cannulation simulator: (a) Side view of simulator; (b) The nine sections of fistula during data collection, used to evenly distribute insertion location.}
     \label{simulator}
\end{figure}

During data collection, the fistula is divided into 9 sections (shown in Fig. \ref{section}), each designated by a two letter code.  The first letter indicates the depth position, (F)ront, (M)iddle, or (B)ack, while the second letter indicates horizontal position, (L)eft, (C)enter, or (R)ight.  For example, ML indicates middle left and BC indicates back center. Balancing insertions between the nine areas helps ensure that the neural network will be able to generalize for any insertion location.

The data collected from the cannulation simulator includes $126$ videos, each of which records a single insertion. For each section of the fistula (See Fig. \ref{section}), there are 14 videos, 10 with infiltration and 4 without infiltration. Next, the videos are split into the test set, the validation set, and the training set. 
The test set and the validation set each have 9 videos with infiltration (one video from each section in Fig. \ref{section}) and 3 videos without infiltration. The training set consists of the remaining 102 videos.

\begin{figure}
    \begin{minipage}[b]{0.3\linewidth}
    \centering
    \centerline{\includegraphics[width=2.4cm, height=1.8cm]{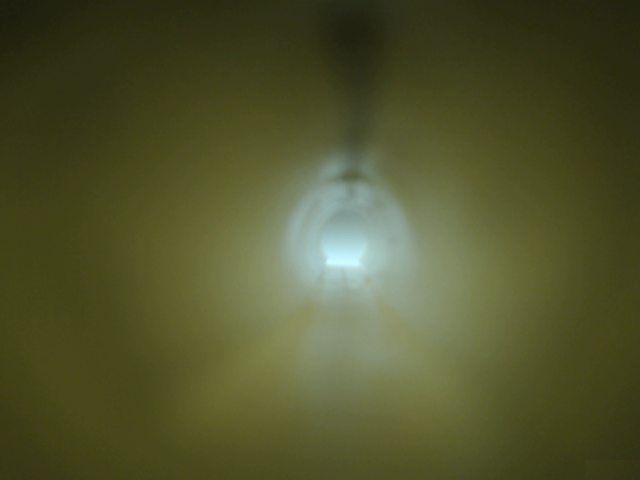}}
    \centerline{\footnotesize{(a) NoNeedle}}\medskip
    \end{minipage}
    \hfill
    \begin{minipage}[b]{0.3\linewidth}
    \centering
    \centerline{\includegraphics[width=2.4cm, height=1.8cm]{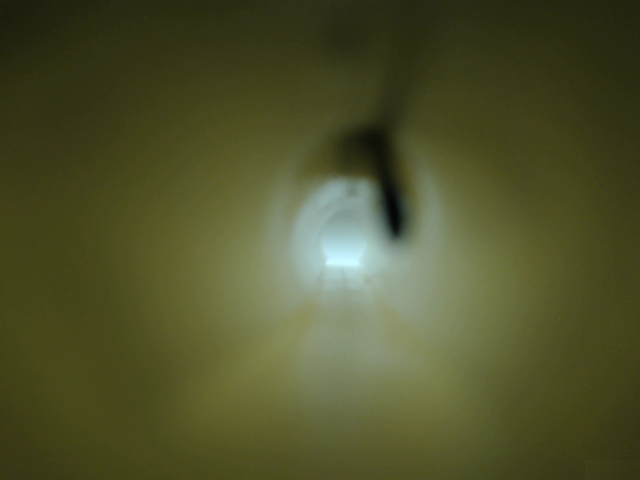}}
    \centerline{\footnotesize{(b) Fist}}\medskip
    \end{minipage}
    \hfill
    \begin{minipage}[b]{0.3\linewidth}
    \centering
    \centerline{\includegraphics[width=2.4cm, height=1.8cm]{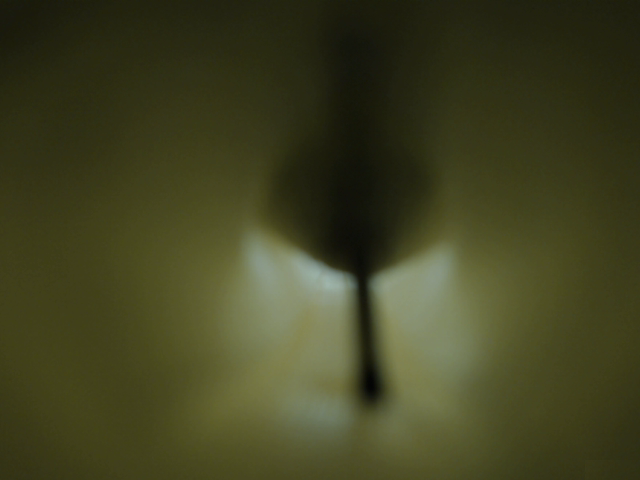}}
    \centerline{\footnotesize{(c) Infil}}\medskip
    \end{minipage}
  \caption{Example images from the three classes in the data set}
  \label{3_classes}
\end{figure}

Since the image frames for the needle insertion are similar to those for the needle removal, the data set consists of three classes: i) NoNeedle - the needle tip is not in the fistula, ii) Fist - the needle tip is in the fistula, and iii) Infil - the needle tip has infiltrated the fistula. For a successful cannulation, the initial video frames would be classified as \qq{NoNeedle} during insertion, then \qq{Fist} while the needle tip is in the fistula, and then \qq{NoNeedle} once the needle tip has left the fistula during removal. For an experiment with infiltration, the sequence of classes would be \qq{NoNeedle},  \qq{Fist},  \qq{Infil}, \qq{Fist}, and finally,  \qq{NoNeedle}. Fig. \ref{3_classes} shows example images from the three classes. 

After extracting frames from the videos, the image frames are hand classified into the 3 classes. Table \ref{img_dist} shows the number of frames from each class for each data set. Note that there is an unbalanced and a balanced version of the validation set.  The balanced validation set has an equal number of frames from each class and is produced by sampling frames from the unbalanced validation set (See Table \ref{img_dist}). The balanced validation set increases the importance of rarely occurring classes, specifically \qq{Fist} in this case. In comparison, the unbalanced validation set preserves the temporal information. As described in Section \ref{training_settings}, the two validation sets were used to train different networks. 

%
\begin{table}
\centering
\caption{The number of images within the data sets}
\label{img_dist}
\begin{tabular}{| c | c | c | c |}
\hline
\rule{0pt}{2ex} 
  Data Set & NoNeedle & Fist & Infil \\
\hline
\rule{0pt}{2ex}  
Training set & 32946 & 8316 & 8700 \\
Test set & 3903 & 1117 & 1170 \\
Unbalanced validation set & 3703 & 970 & 1137\\
Balanced validation set & 970 & 970 & 970 \\
\hline
\end{tabular}
\end{table}

\subsection{Modified pre-trained models}

The first two networks examined in this work are modified versions of VGG-16 \cite{VGG16} and ResNet-50 \cite{ResNet50}. These networks achieved state-of-the-art performance in ImageNet competition and so  are commonly used with transfer learning as benchmarks for comparison. 

\begin{figure}[!t]
  \centering
  \includegraphics[scale=0.4]{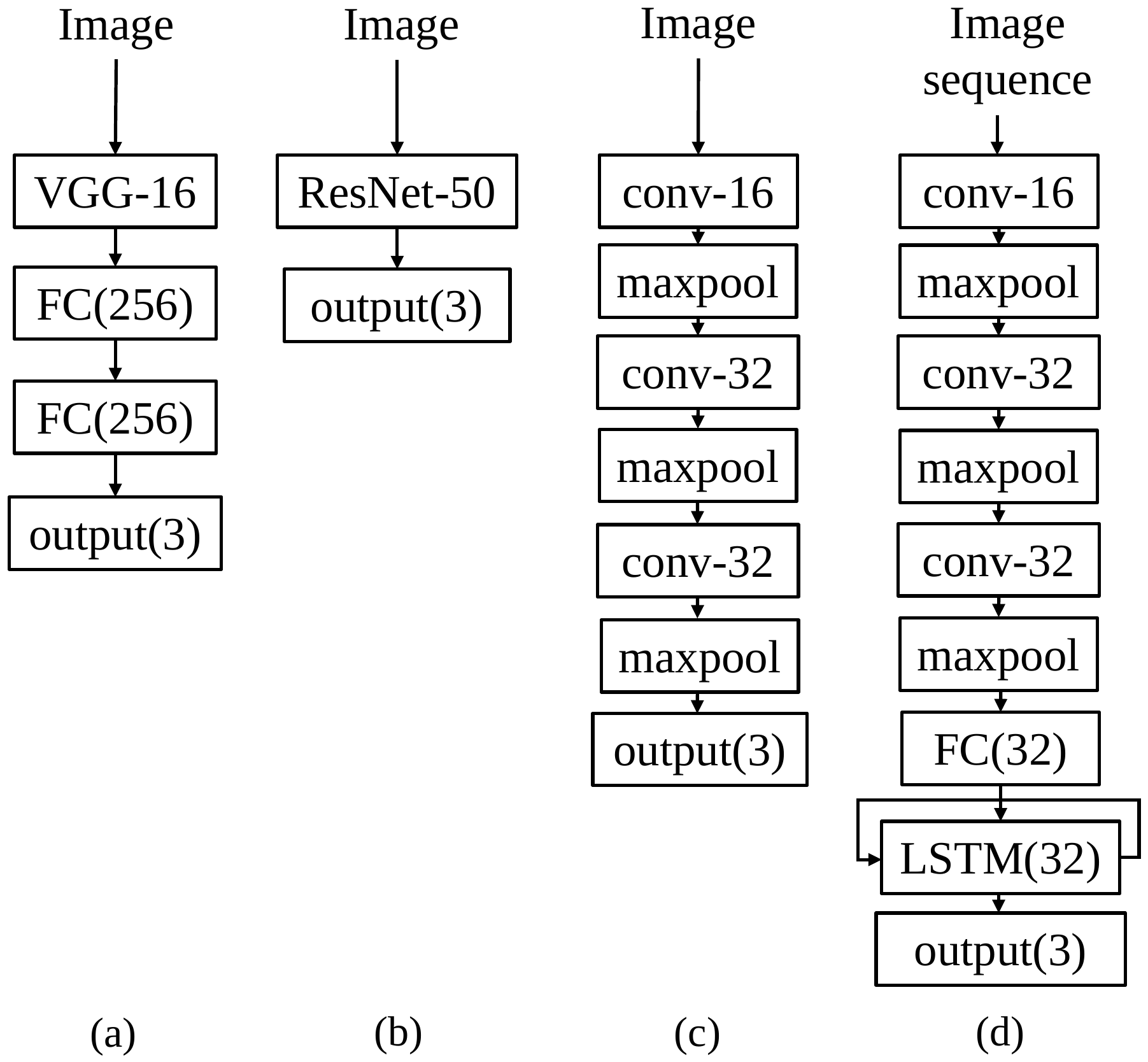} 
  \caption{Network architecture: (a) Modified VGG-16; (b) Modified ResNet-50; (c) Light CNN with three 2-layer blocks; (d) CRNN. The number within parentheses denotes the number of neurons within a specific layer, while \lq conv-$\alpha$' denotes a convolutional layer with $\alpha$ filters}
  \label{networks}
\end{figure}

\noindent \textbf{VGG-16:} The VGG-16 network includes a stack of convolutional layers and max-pooling layers, following three fully-connected (FC) layers. To utilize the VGG-16 network, the three FC layers are shrunk. Specifically, each of the two hidden FC layers only has 256 neurons, while the output layer has 3 neurons corresponding to the 3 classes. By shrinking the size of the FC layers, the number of parameters within the VGG-16 network decreases from about 138M to about 21M. The two hidden FC layers and the output layer use ReLU activation functions and softmax activation functions, respectively. The architecture of the modified VGG-16 is shown in Fig. \ref{networks}a. To avoid overfitting, we also trained a modified VGG-16 model with a dropout layer (the dropout ratio is 0.5) between each two FC layers. Hereafter the network with dropout layers is abbreviated as VGG-16(w/do). When training the two VGG-16 networks, we applied transfer learning and only updated the parameters within the 3 FC layers.

\noindent \textbf{ResNet-50:} The ResNet-50 network does not have any hidden FC layers. To adopt ResNet-50, the number of outputs is reduced from 1000 to 3, corresponding to our smaller number of classes. During the training process, we utilized transfer learning and only updated parameters within the output layer. The modified ResNet-50 is shown in Fig. \ref{networks}b. 

Before training VGG-16 and ResNet-50 on the present data set, the images were modified based on the requirements of the pre-trained models. Specifically, the pixel values were zero-centered and the images were resized to $224\times224\times3$. During the training process, the images were shuffled after each epoch. 


\subsection{Light convolutional neural networks (light CNNs)}\label{light_CNN}

Compared with the 1000-class ImageNet data set, our data set only has 3 classes, which suggests that a smaller network should be  capable of this smaller classification task. Inspired by the structure of VGG-16, we propose ``light CNNs'' that use $3\!\times3\!$ convolutional filters but with only two or three convolutional layers, and hence significantly less parameters.

The light CNN architecture includes a stack of 2-layer blocks. Each block consists of a convolutional layer followed by a max-pooling layer. The convolutional layer has $3\!\times3\!$ filters with stride 1. The ReLU activation function is used by the convolutional layer. To preserve the spatial resolution, the padding is set to 1 pixel. The max-pooling layer uses $2\times2$ pooling windows with stride 2.

To determine the best architecture of our light CNN, we investigated a variety of 2-block and 3-block light CNNs with different numbers of convolutional filters. Table \ref{result_lcnn} lists the examined networks, where the network architecture is denoted by the number of convolutional filters in each CNN block, from input to output. For example, the $\langle 8\!-\!16\rangle$ CNN has 8 convolutional filters in the first block and 16 in the second block.  


Fig. \ref{networks}c shows the structure of the $\langle 16\!-\!32\!-\!32\rangle$ CNN, which consists of three blocks with corresponding convolutional layers consisting of 16, 32, and 32 filters.  Since the feature map size becomes halved after passing through a max-pooling layer, the number of filters within the subsequent convolutional layer is preserved or increased so as to preserve the complexity. 

The output layer of our light CNNs has 3 neurons with softmax activation functions. Similar to ResNet-50, our light model does not have hidden FC layers. FC layers increase the possibility of overfitting, since they contain so many more trainable parameters.

The data preprocessing method includes resizing images and zero-centering. The target size of the images is $112\times112\times3$. We shrink the dimension of input images from $224\times224\times3$ to $112\times112\times3$ to reduce computational cost without affecting the performance of the light CNNs. The image pixels are zero-centered by subtracting the mean RGB values computed on the ImageNet data set. When training the light CNNs, the images are shuffled after each epoch.

\subsection{Convolutional recurrent neural networks (CRNNs)}
Our light CNN models generate output based on a single image, ignoring temporal information. The classification problem, however, involves video frames for which temporal information is important. For instance, an image labelled as \qq{Infil} cannot follow that with the \qq{NoNeedle} label. Assuming the temporal information is beneficial to our task, we construct the CRNN models by adding a LSTM layer to our best light CNN (shown in Fig. \ref{networks}c).


To take advantage of the trained light CNN model, the CRNN is formed by adding a 32-neuron FC layer with ReLU activation functions onto the  last max-pooling layer of the light CNN. Then, a 32-neuron LSTM layer is attached to the FC layer, followed by a 3-neuron output layer with softmax activation functions. When training the CRNNs, the parameters within the convolutional layers were frozen. To avoid overfitting, half of the LSTM neurons were randomly dropped (i.e., the dropout ratio is 0.5) during the training process. The structure of our CRNN is shown in Fig. \ref{networks}d.

To determine the best CRNN model, we trained a variety of CRNNs with different time steps. The CRNN model with the highest validation accuracy was then chosen as our best CRNN.

Since our CRNN model includes a pre-trained light CNN, the preprocessing method described in Section \ref{light_CNN} is used. The sequential images, however, are fed to the network without shuffling and data augmentation. 

\subsection{Training settings}\label{training_settings}
During the training process, the Adam optimizer \cite{Adam} was used by all the networks. The number of epochs and the learning rate were set to $50$ and $0.0001$, respectively, chosen empirically. When training each network, we only saved the parameters corresponding to the highest validation accuracy. 

To increase variation in the training images, the following data augmentation techniques were applied to the training set:
\begin{itemize}
    \item Random horizontal shift, less than 20\% image width.
    \item Random vertical shift, less than 20\% image height. 
    \item Random rotation in range $[-10^{\circ}, 10^{\circ}]$.
    \item Random shift in brightness in range $[0.5, 1.5]$.
\end{itemize}
Since data augmentation affects the temporal information, these techniques were only used when training the modified pre-trained models and the light CNNs.

Since an unbalanced training set was used (See Table \ref{img_dist}), class weights were applied while  training the networks, specifically, 1.0, 3.96, and 3.79 for \qq{NoNeedle},  \qq{Fist}, and \qq{Infil}, respectively. These weights are inversely proportional to the number of training examples from the corresponding class. This balances the importance of each class when training the CNNs. For the CRNNs, the class weights increase the accuracy of predicting the minority class (i.e., \qq{Fist}) even though the training and validation sets are unbalanced.  

Two validation sets (shown in Table \ref{img_dist}) were adopted during the training process. Specifically, the balanced validation set was used for training the modified pre-trained models and the light CNNs. Since sampling distorts the image sequence, the unbalanced validation set was applied when training the CRNNs.

\subsection{Real-time experiments}\label{rt_experiment}

To test if the light CNN and the CRNN are suitable for use in the cannulation simulator, a commodity PC configured with the comparatively large network (i.e., the CRNN) was used to analyze real-time video streams from the camera in the simulator. Since the CRNN is larger than the light CNN, if the CRNN can be implemented in real time, then the smaller light CNN can also be implemented in real time. The PC is equipped with Intel i5-7300HQ CPU and Nvidia GTX-1050 GPU. The software is written on Keras 2.1.5 with TensorFlow backend.

The real-time experiments included 36 insertions in total, specifically four insertions at each of the nine sections shown in Fig. \ref{section}. For each section, two insertions had infiltration.

\section{Results and Discussion}
The validation and test data sets each consist of 12 videos, and each video records a single insertion.  In this paper, "validation accuracy"  and "test accuracy"  refer to the average accuracy on the validation data set and test data set, respectively.  Validation accuracy is used to determine the optimal network structure, and the test accuracy is used to evaluate the performance of the optimal network.
\subsection{Results of the pre-trained models} 
Table \ref{result_pt} shows the number of parameters, the validation accuracy, and the test accuracy for the two modified VGG-16 networks and the modified ResNet-50. The two VGG-16 networks achieve similar performance on both the validation set and the test set. Among the three modified pre-trained models, the ResNet-50 achieves the highest test accuracy (0.971). Note that the three networks have similar numbers of parameters (about 22M).
\begin{table}
\centering
\caption{A comparison of pre-trained models}
\label{result_pt}
\begin{tabular}{| c | c | c | c |}
\hline
\rule{0pt}{2ex} 
  & VGG-16 & VGG-16(w/do) & ResNet-50 \\
\hline
\rule{0pt}{2ex}  
\# param. & 21M & 21M & 23M \\
val. acc. & 0.969 & 0.967 & 0.961 \\
test acc. & 0.967 & 0.968 & 0.971\\
\hline
\end{tabular}
\end{table}

\subsection{Results of light CNNs}

Table \ref{result_lcnn} compares the number of parameters and the validation accuracy for different light CNNs. The  $\langle 16\!-\!32 \rangle$ CNN has the highest validation accuracy (0.985) among the 2-block light CNNs. The $\langle 16\!-\!32\!-\!32 \rangle$ CNN adds a third block to the $\langle 16\!-\!32 \rangle$ CNN, which improves the validation accuracy to 0.988. Consequently, the $\langle 16\!-\!32\!-\!32 \rangle$ CNN is our best light CNN.


\begin{table*}[t]
\centering
\caption{A comparison of light CNNs}
\label{result_lcnn}
\begin{tabular}{| c || c | c | c | c | c | c | c || c | c |}
\hline
\rule{0pt}{2ex} 
Architecture & $\langle2\!-\!4\rangle$ & $\langle4\!-\!8\rangle$ & $\langle8\!-\!16\rangle$ & $\langle16\!-\!32\rangle$ & $\langle32\!-\!64\rangle$ & $\langle64\!-\!128\rangle$ & $\langle128\!-\!256\rangle$ & $\langle8\!-\!16\!-\!32\rangle$ & $\langle16\!-\!32\!-\!32\rangle$ \\
\hline
\rule{0pt}{2ex}  
\# param. & 10K & 19K & 39K & 80K & 170K & 377K & 901K & 25K & 33K \\
val. acc. & 0.927 & 0.967 & 0.978 & \textbf{0.985} & 0.980 & 0.981 & 0.983 & 0.983 & \textbf{0.988} \\
\hline
\end{tabular}
\end{table*}

Compared with the modified pre-trained models (shown in Table \ref{result_pt}), the $\langle 16\!-\!32\!-\!32 \rangle$ CNN has higher validation accuracy (0.988). Also, the test accuracy for $\langle 16\!-\!32\!-\!32 \rangle$ CNN is 0.983, which is higher than that for our modified ResNet-50 (0.971). It should be noted that the $\langle 16\!-\!32\!-\!32 \rangle$ CNN has only 33K parameters, which is about 0.2$\%$ of the parameters of the modified VGG-16 (about 21M).

\subsection{Results of CRNNs}

\begin{figure}[!t]
  \centering
  \includegraphics[scale=0.5]{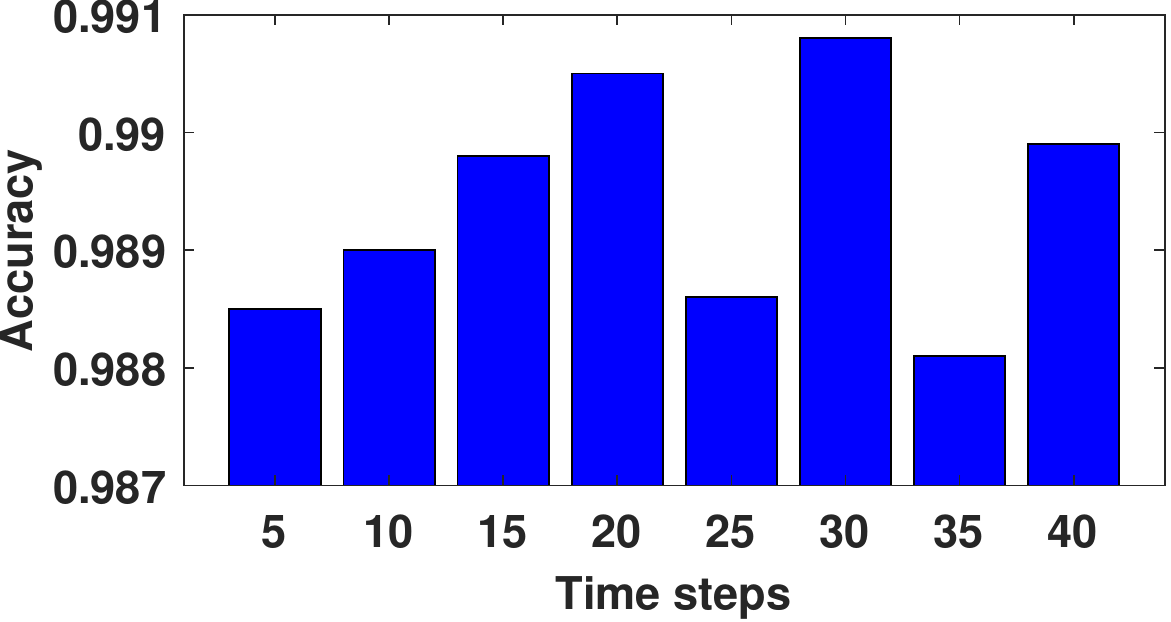}
  \caption{The validation accuracy for CRNNs with different time steps}
  \label{CRNN_acc}
\end{figure}

Fig. \ref{CRNN_acc} shows the validation accuracy for CRNNs with different time steps. Since time steps only affect the cell state within the LSTM layer, all our CRNN models have identical number of parameters (about 223K), which is about 1$\%$ as small as that for the modified VGG-16.

As shown in Fig. \ref{CRNN_acc}, our CRNN model achieves the highest validation accuracy (0.9908) when time steps are 30. The CRNN with 30 time steps (abbreviated as CRNN(30ts) hereafter) is then chosen as our best CRNN model. Although CRNN(30ts) outperforms the CRNN models with other time steps, all of the examined CRNNs have similar validation accuracy. Among all of our networks, CRNN(30ts) has the highest validation accuracy. The test accuracy of CRNN(30ts) is identical to the test accuracy of the $\langle 16\!-\!32\!-\!32 \rangle$ CNN.

\subsection{Comparison of the test accuracy}
\begin{figure}
    \hspace{0.2cm}
    \begin{minipage}[b]{0.4\linewidth}
    \centering
    \centerline{\includegraphics[width=3.5cm, height=3.5cm]{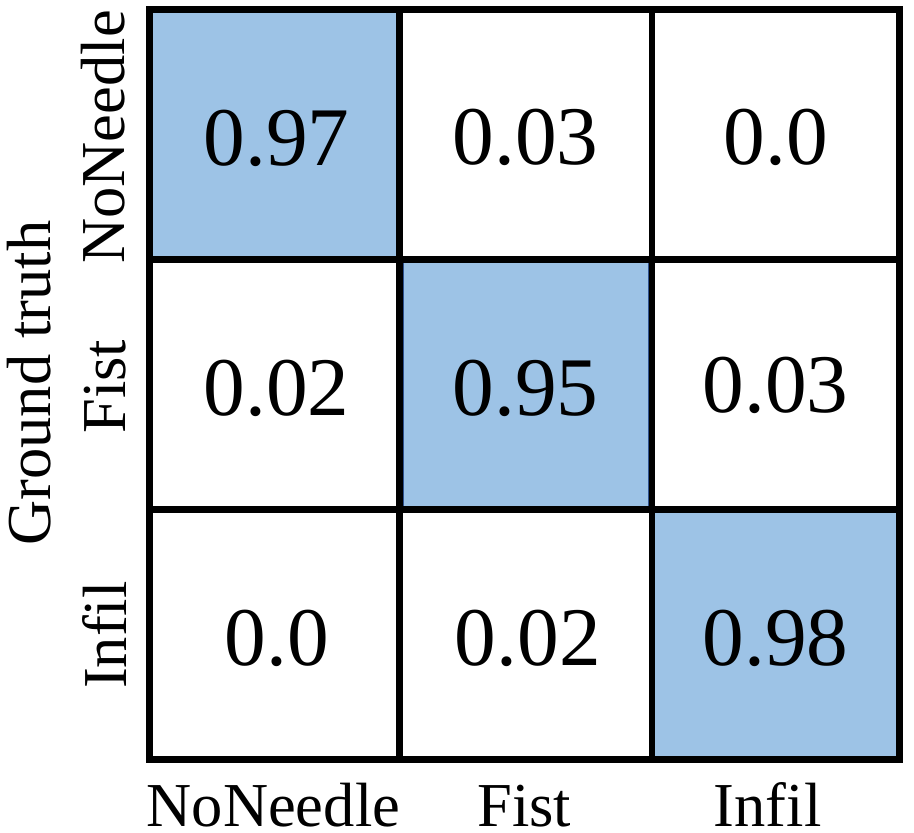}}
    \centerline{\footnotesize{(a) VGG-16(w/do), test acc.=0.968}}\medskip
    \end{minipage}
    \hspace{0.8cm}
    \begin{minipage}[b]{0.4\linewidth}
    \centering
    \centerline{\includegraphics[width=3.5cm, height=3.5cm]{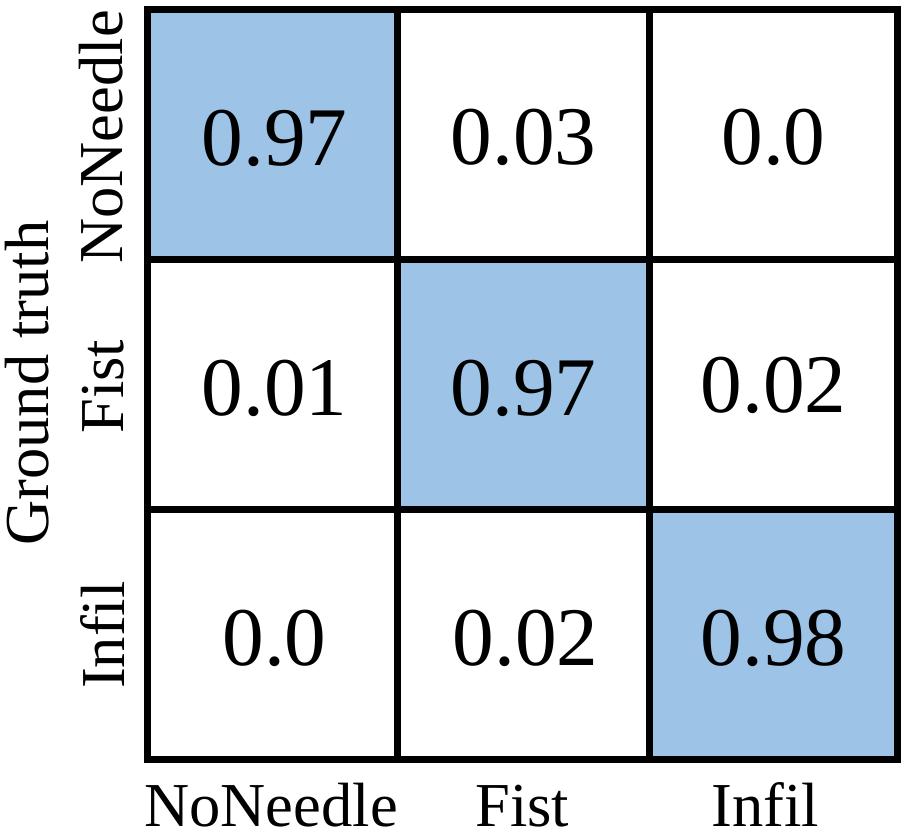}}
    \centerline{\footnotesize{(b) ResNet-50, test acc.=0.971}}\medskip
    \end{minipage}
    
    \hspace{0.2cm}
    \begin{minipage}[b]{0.4\linewidth}
    \centering
    \centerline{\includegraphics[width=3.5cm, height=3.5cm]{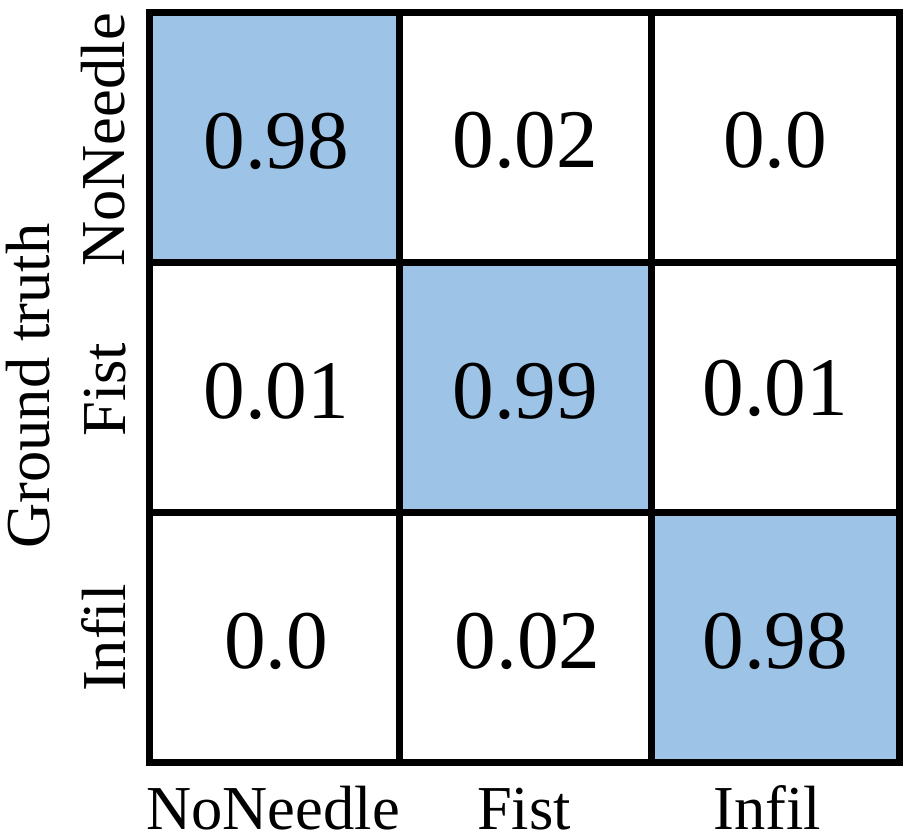}}
    \centerline{\footnotesize{(c) $\langle 16\!-\!32\!-\!32\rangle$, test acc.=0.983}}\medskip
    \end{minipage}
    \hspace{0.8cm}
    \begin{minipage}[b]{0.4\linewidth}
    \centering
    \centerline{\includegraphics[width=3.5cm, height=3.5cm]{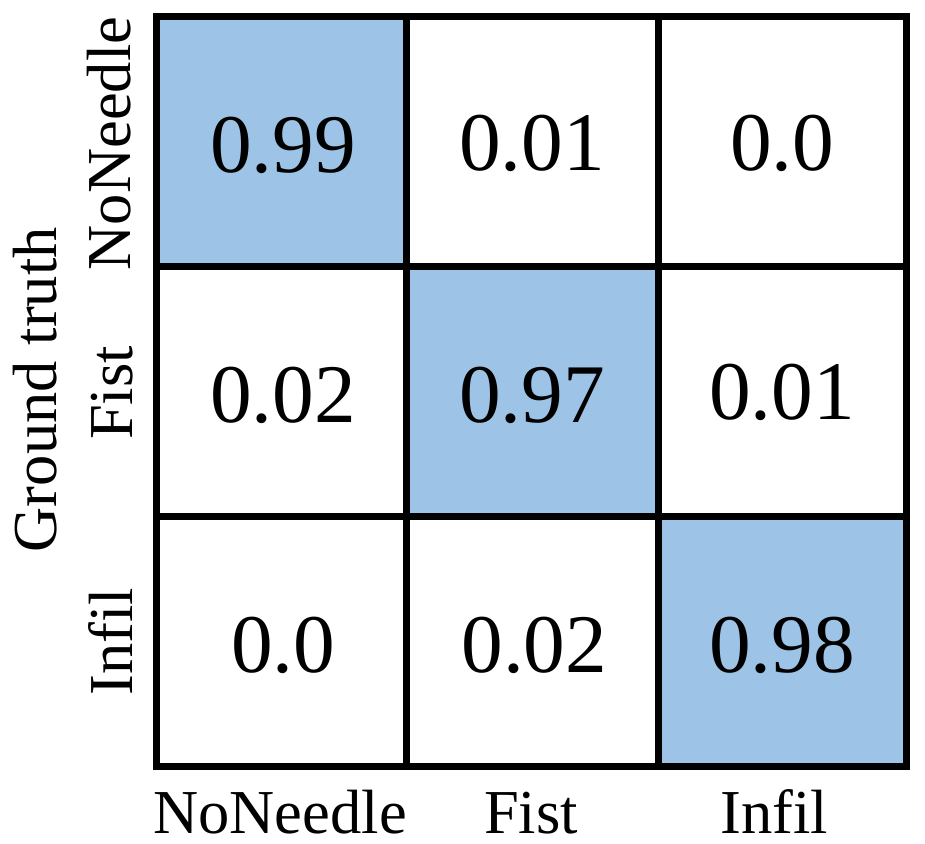}}
    \centerline{\footnotesize{(d) CRNN(30ts), test acc.=0.983}}\medskip
    \end{minipage}
  \caption{The confusion matrices for the test set for four different neural networks. Note that the second row within (c) has a round-off error}
  \label{confusion_mat_p}
\end{figure}

To compare the network performance, the VGG-16(w/do), the modified ResNet-50, the $\langle 16\!-\!32\!-\!32 \rangle$ CNN, and CRNN(30ts) are used to process the test set and generate confusion matrices (shown in Fig. \ref{confusion_mat_p}).  

The confusion matrix for the VGG-16(w/do) and for the modified ResNet-50 are shown in Fig. \ref{confusion_mat_p}a and \ref{confusion_mat_p}b, respectively. Both pre-trained models achieve the highest accuracy when detecting \qq{Infil}. Compared with the VGG-16(w/do), the modified ResNet-50 achieves similar detection accuracy for each class.

The confusion matrix for the $\langle 16\!-\!32\!-\!32 \rangle$ CNN is shown in Fig. \ref{confusion_mat_p}c. Compared with Fig. \ref{confusion_mat_p}b, the $\langle 16\!-\!32\!-\!32 \rangle$ CNN has higher detection accuracy for the class \lq NoNeedle' and \lq Fist'. Note that the sum of accuracy for ground truth \lq Fist' is not equal to 1 due to the round-off error.

Fig. \ref{confusion_mat_p}d demonstrates the confusion matrix for CRNN(30ts). Among our networks, CRNN(30ts) achieves the highest test accuracy for \qq{NoNeedle}.

\begin{figure}[!t]
  \centering
  \includegraphics[scale=0.6]{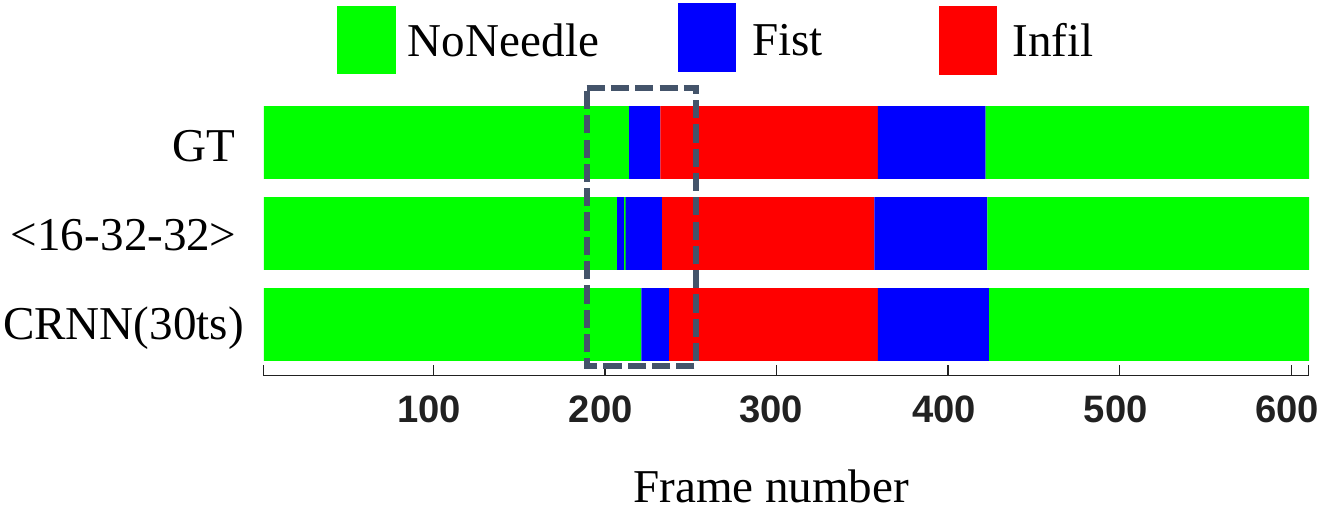} 
  \caption{Comparison of the processing results of $\langle 16\!-\!32\!-\!32 \rangle$ CNN, CRNN(30ts), and ground truth (GT). The gray rectangle denotes the period with sudden switches. The processed video includes infiltration}
  \label{v_infil}
\end{figure}

\begin{figure}[!t]
  \centering
  \includegraphics[scale=0.6]{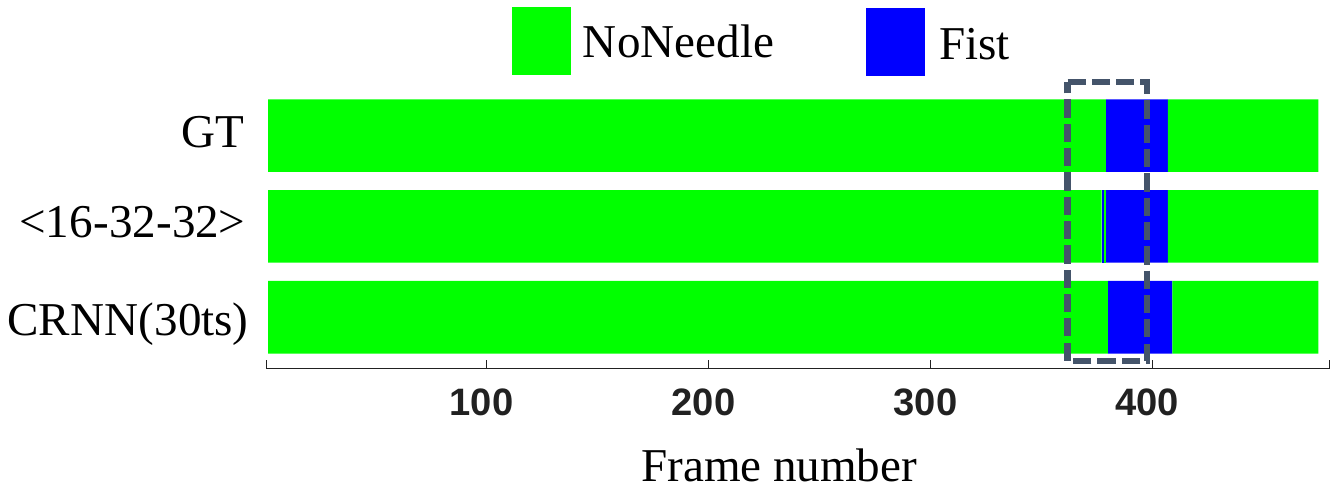} 
  \caption{Comparison of the processing results of $\langle 16\!-\!32\!-\!32 \rangle$ CNN, CRNN(30ts), and ground truth (GT). The gray rectangle denotes the period with sudden switches. The processed video is free of infiltration}
  \label{v_no_infil}
\end{figure}

To illustrate the effectiveness of the LSTM layer, the output of the $\langle 16\!-\!32\!-\!32 \rangle$ CNN and CRNN(30ts) are visualized for two videos from the test set, one with infiltration  (shown in Fig. \ref{v_infil}) and one without infiltration (shown in Fig. \ref{v_no_infil}). As can be seen from Fig. \ref{v_infil} and \ref{v_no_infil}, the detection results of the $\langle 16\!-\!32\!-\!32 \rangle$ CNN rapidly switch between classes at the beginning of \lq Fist', while CRNN(30ts) does not show such oscillatory behavior. 

\subsection{Results of real-time experiments}
Before attempting the real-time experiments, we used the PC (described in Section \ref{rt_experiment}) to examine the processing speed of CRNN(30ts). Initial results show that CRNN(30ts) can analyze up to 54 frames per second (fps), which is faster than the 30 fps video stream from the camera. 

 For the 36 real-time experiments, the maximum accuracy (acc$_{max}$) is 0.997, the minimum accuracy (acc$_{min}$) is 0.955, and the average accuracy (acc$_{avg}$) is 0.987.  To identify whether classification accuracy depends on insertion location, the 9 sections shown in Fig. \ref{section} are classified into 6 groups, namely Left, Center, Right, Front, Middle, and Back. An BL insertion would be included in both the Back and Left groups. Each group includes 12 insertions. Table \ref{result_rt} summarizes the real-time detection accuracy of CRNN(30ts). 

As shown in Table \ref{result_rt}, CRNN(30ts) has similar performance at Left, Center, and Right. For the group Front, Middle, and Back, the CRNN achieves the highest average accuracy at Front, followed by Middle and then Back. This suggests that the network detection accuracy decreases as the distance between the camera and the needle increases.

 
 As can be seen from Table \ref{result_rt}, CRNN(30ts) achieves similar acc$_{avg}$ for the experiments with infiltration and the experiments without infiltration (Or more specifically, the Wilcoxon rank sum test indicates that the medians are not significantly different with $p=0.51$). Note that there is an outlier (acc$_{min}$=0.955) in the experiments with infiltration.
 
\begin{table*}[t]
\centering
\caption{Detection accuracy for the 36 real-time experiments}
\label{result_rt}
\begin{tabular}{| c || c | c | c || c | c | c || c | c |}
\hline
\rule{0pt}{2ex} 
Group & Left & Center & Right & Front & Middle & Back & No infiltration & Infiltration \\
\hline
\rule{0pt}{2ex}  
\# trials & 12 & 12 & 12 & 12 & 12 & 12 & 18 & 18 \\
\hline
acc$_{max}$ & 0.997 & 0.997 & 0.997 & 0.997 & 0.997 & 0.995 & 0.997 & 0.997 \\
acc$_{min}$ & 0.955 & 0.974 & 0.976 & 0.981 & 0.976 & 0.955 & 0.976 & 0.955 \\
acc$_{avg}$ & 0.984 & 0.989 & 0.987 & 0.990  & 0.987 & 0.983 & 0.988 & 0.985 \\

\hline
\end{tabular}
\end{table*}

\begin{figure}[!htbp]
  \centering
  \includegraphics[scale=0.6]{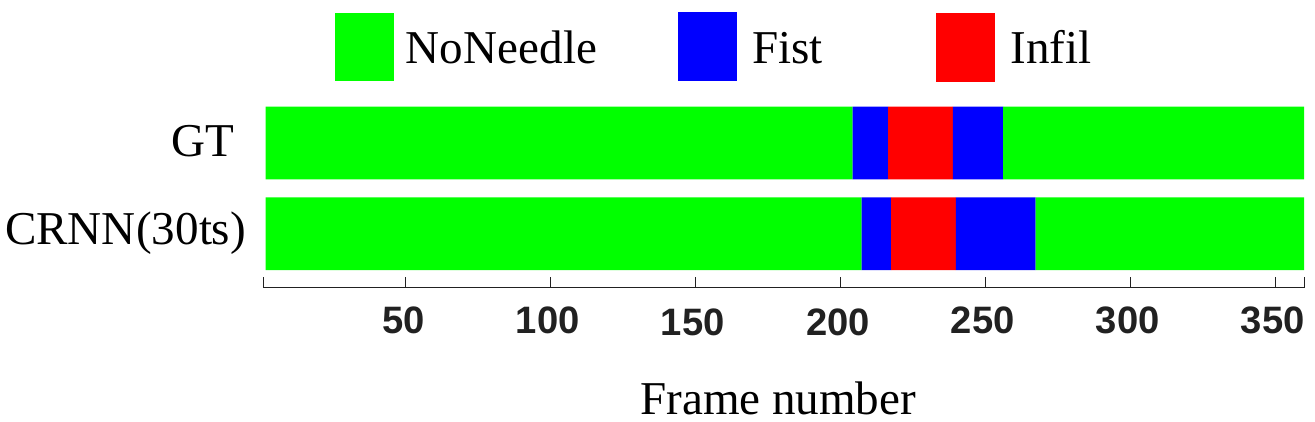} 
  \caption{Comparison between the real-time detection results of CRNN(30ts) and ground truth (GT). The processed video has the lowest detection accuracy among the 36 real-time experiments}
  \label{v_30_rt}
\end{figure}

Fig. \ref{v_30_rt} shows the detection results of the real-time experiment with the lowest accuracy (0.955). This insertion was performed at section BL. As can be seen from  Fig. \ref{v_30_rt}, the detection results of CRNN(30ts) lag behind the ground truth, especially as in the switch from `Fist' to `NoNeedle'.  The needle tip appears less distinct when farther from the camera, and this may be the reason for the lag. Detection accuracy for these more difficult sections could be improved by increasing the number of training examples from these sections. The classification errors tend to occur close in time to transitions in the ground truth needle state. For the 36 real-time experiments, the maximum difference between the predicted transition and the ground truth is 11 frames, while the average difference between the predicted transition and the ground truth is 1.36 frames. These sorts of error will not significantly impact the use of the needle state information in the cannulation simulator.

\section{CONCLUSIONS}

In this study, needle puncture and infiltration are detected in a cannulation simulator using video analysis with artificial neural networks. Of the four different kinds of artificial neural networks examined, the experimental results show that the  $\langle 16\!-\!32\!-\!32 \rangle$ CNN and CRNN(30ts) have the best performance among the networks. Thanks to the small size of the CRNN, the network can run on inexpensive hardware and process real-time video streams. Consequently, our CRNN-based system provides a solution for basic cannulation training using affordable simulators.

Our system still has some limitations. First, our data set only includes videos with bottom wall infiltrations, so the detection accuracy for side wall infiltrations is not guaranteed. Second, our camera requires an artificial fistula with a large diameter. Furthermore, the video-based methods used here require a straight fistula to provide a clear view. In the future, we will consider side wall infiltrations when creating the data set, and we expect to see similar performance. Also, we will evaluate cameras suitable for artificial fistulas with small diameters. Moreover, we will use the network output to simulate blood flashback and trigger warnings when infiltration occurs. We also plan to gather user experience data from nursing students working with the simulator.

\addtolength{\textheight}{-12cm}   




\section*{ACKNOWLEDGMENT}

Research reported in this publication was supported by a NIH/NIDDK K01 Award (K01DK111767).
Clemson University is acknowledged for generous allotment of compute time on Palmetto cluster.

\bibliographystyle{IEEEtran}
\typeout{}
\bibliography{reference}

\end{document}